\documentclass[conference]{IEEEtran}
\IEEEoverridecommandlockouts
\usepackage{cite}
\usepackage{hyperref}
\usepackage{amsmath,amssymb,amsfonts}
\usepackage{algorithm}
\usepackage{algorithmic}

\usepackage{graphicx}
\usepackage{textcomp}
\usepackage{xcolor}
\usepackage{booktabs}
\usepackage{multirow}
\usepackage{threeparttable}
\def\BibTeX{{\rm B\kern-.05em{\sc i\kern-.025em b}\kern-.08em
    T\kern-.1667em\lower.7ex\hbox{E}\kern-.125emX}}
\begin{document}

\title{CipherFormer: Efficient Transformer Private Inference with Low Round Complexity}

\author{\IEEEauthorblockN{Weize Wang\IEEEauthorrefmark{1}, Yi Kuang\IEEEauthorrefmark{1}}
\IEEEauthorblockA{\IEEEauthorrefmark{1} School of Cyber Science and Engineering, Shanghai Jiao Tong University, Shanghai, China}
\IEEEauthorblockA{\{wang-weize, schemer\}@sjtu.edu.cn}
}

\maketitle

\begin{abstract}
There is a growing trend to outsource the inference task of large transformer models to cloud servers. However, this poses a severe threat to users' private data as they are exposed to cloud servers after uploading. Although several works attempted to provide private inference for transformer models, their hundreds of communication rounds limit the application scenarios. Motivated by the desire to minimize round complexity, we propose CipherFormer, a novel transformer private inference scheme using homomorphic encryption and garbled circuits. We present a protocol for quickly computing homomorphic matrix multiplications. We then modify the attention mechanism and design the corresponding garbled circuits. Furthermore, we show how to use a lightweight attention mechanism and mixed-bitwidth to reduce the inference latency while maintaining accuracy. In comparison with an advanced homomorphic encryption scheme on text classification tasks, our model improves accuracy by 3\% to 11\% while performing private inference with a 7.7x-11.9x speedup.
\end{abstract}


\section{Introduction}

Transformer models, proposed by \cite{DBLP:conf/nips/VaswaniSPUJGKP17}, are highly successful in diverse domains such as Computer Vision (CV) and Natural Language Processing (NLP). A notable example is the Generative Pre-trained Transformer (GPT) \cite{radford2018improving}, a leading language model recognized for its impressive performance. Despite GPT's capabilities, its large model size poses challenges for independent training, leading many users to resort to outsourcing data processing to cloud service providers. In this scenario, the cloud server handles model inference, enabling users to await results. However, this outsourced approach raises privacy concerns, especially in collaborative systems with a cloud-based GPT, where untrusted servers may compromise data confidentiality. Recognizing the unrealistic expectation of cloud providers disclosing model parameters, it's crucial to carefully address potential data exposure risks in collaborative environments.

\begin{figure}[hb]
\centerline{\includegraphics[width=0.4\textwidth]{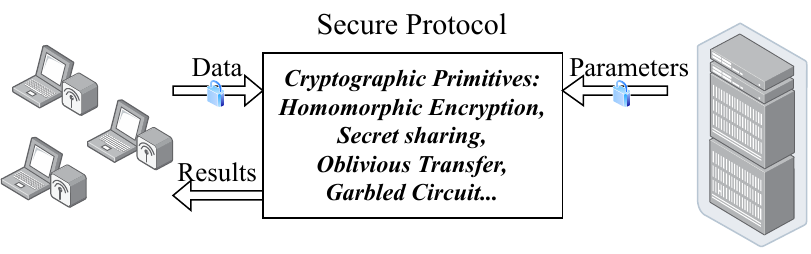}}
\caption{Private model inference between clients and servers.}
\label{secureInfer}
\end{figure}

Fortunately, cryptographic primitives like homomorphic encryption (HE), secret sharing (SS), oblivious transfers (OT), and garbled circuits \cite{DBLP:conf/focs/Yao86} (GC) address privacy concerns in model inference. In outsourced computation scenarios (Figure~\ref{secureInfer}), the clients and server participate in secure protocols using cryptographic primitives. The clients input data, and the server provides model parameters for inference. Following the protocol, the clients receive inference results, ensuring that neither the data nor the model parameters are revealed to each other.

Several related works \cite{DBLP:conf/nips/HaoLCXXZ22,DBLP:journals/corr/abs-2211-01452,Zeng_2023_ICCV,DBLP:conf/dac/ZhengLJ23,DBLP:journals/corr/abs-2308-09923} have been dedicated to crafting private inference protocols for transformer models. These protocols primarily leverage SS and OT, leading to a high number of communication rounds, often reaching into the hundreds. This necessitates active client involvement in frequent communication with the server throughout the entire inference process. Such a demand significantly increases the time cost, particularly in unstable network environments.

In contrast, other works based on HE \cite{chen-etal-2022-x,comi2021herbert} provide a partial resolution to this challenge by enabling the non-interactive inference of multiplication and addition over ciphertext. However, there are numerous operations in transformer models, such as $Softmax$, that cannot be represented using only multiplication and addition. As a result, these approaches suffer from low accuracy.

To address the dilemma, the classic approach Gazelle \cite{DBLP:conf/uss/JuvekarVC18} proposes a hybrid method that combines Homomorphic Encryption (HE) with Garbled Circuits (GC), characterized by a constant round complexity, thereby reducing communication rounds. However, while Gazelle is effective for fully connected networks and convolutional neural networks \cite{DBLP:conf/nips/KrizhevskySH12}, it faces limitations when applied to more complex transformer models.

We propose CipherFormer, a novel approach that combines HE and GC to facilitate private inference for transformer models. We indicate that Gazelle is incapable of solving two crucial operations in transformer models: the multiplication between two ciphertext matrices, and the complex $Softmax$ activation function in the attention mechanism. To address these challenges, we first subtly convert the ciphertext matrix multiplication into a faster public matrix multiplication with lower noise growth, and the designed multiplication sub-protocol exclusively requires 2 communication rounds. Secondly, we simplify the $Softmax$ function within the attention mechanism and devise an customized GC construction for it.

Furthermore, we devise two optimization strategies aimed at reducing the latency of CipherFormer inference while maintaining inference accuracy. The initial optimization introduces a lightweight attention mechanism, reducing the computational complexity associated with ciphertext multiplication and minimizing the communication overhead of the activation function in GC. The second optimization draws on research \cite{DBLP:conf/sp/RatheeRGGSCR21} and introduces mixed bitwidth into the protocol to further reduce communication overhead.


Experiments conducted on several text categorization benchmarks reveal the superior performance of our scheme in terms of both accuracy and latency. When compared to the advanced approach HErBERT \cite{comi2021herbert}, our schemes demonstrate accuracy improvement in private inference ranging from 3\% to 11\%, coupled with a significant speedup ranging from 7.7x to 11.9x.

\section{Preliminaries}

\subsection{Threat Model}
Our threat model aligns with prior work, specifically SecureML \cite{DBLP:conf/sp/MohasselZ17} and Gazelle \cite{DBLP:conf/uss/JuvekarVC18}. In this model, a client possesses private data, while a server holds a trained transformer model. Both the server and client are considered semi-honest, meaning that each party seeks to access the other's private information while correctly following the protocol. Importantly, our protocol ensures that neither the client's input data nor the server's model weight information can be obtained by the other party.

\subsection{Transformer and Attention mechanism}
The Transformer, introduced by \cite{DBLP:conf/nips/VaswaniSPUJGKP17}, revolutionized NLP and other sequential data processing tasks. It consists of an encoder-decoder architecture, each with a similar structure. Unlike traditional sequence models like recurrent neural networks \cite{DBLP:journals/cogsci/Elman90}, the Transformer relies on the attention mechanism. This mechanism permits the model to assign varying weights (i.e. attention) to different segments of the input sequence. The attention mechanism is particularly advantageous for capturing long-range dependencies, rendering the Transformer highly effective in applications such as language translation and text generation. However, the intricate nature of attention, involving a large multiplication depth and complex operations such as $Softmax$, poses challenges in the design of efficient and accurate private inference protocols. In this study, we narrow our focus to text categorization tasks, aligning with prior work \cite{DBLP:conf/nips/HaoLCXXZ22}. This specific task allows us to concentrate solely on the encoder part when formulating the privacy protocol.

\subsection{Cryptographic Primitives}
Our protocol design aligns with the approach adopted by Gazelle, wherein HE is employed for the linear component of the model, while GC is utilized for the non-linear operations. The switching between these cryptographic primitives is facilitated by a straightforward additive secret sharing protocol proposed in Gazelle. To execute basic HE operations including addition, scalar multiplication ($\Pi_{SIMDScMult}$), and slot permutation, we utilize packed additively homomorphic encryption (PAHE) \cite{DBLP:conf/uss/JuvekarVC18}, which is implemented by the BFV  \cite{DBLP:conf/crypto/Brakerski12, DBLP:journals/iacr/FanV12} algorithm.

We adopt 20-bit fixed-point numbers \cite{DBLP:conf/fc/CatrinaS10} with a 9-bit fractional part bitwidth to represent both the client input and model parameters. This choice is motivated by the fact that the plaintext space of BFV algorithm is defined by the ring $R_{p}^{n}:=\mathbb{Z}_{p}[x] /\left(x^{n}+1\right)$, where $p$ is a 20-bit prime.

\section{CipherFormer}
In this section, we begin with an analysis to determine suitable cryptographic primitives for each layer of the transformer model, and indicate that previous work is failing to handle ciphertext matrix multiplication and activation in attention mechanism. Consequently, we introduce customized algorithms to address these specific operations in the context of privacy-preserving inference. Following this, we present two optimization strategies aimed at balancing inference accuracy and latency.

\begin{figure}[htbp]
\centerline{\includegraphics[width=0.4\textwidth]{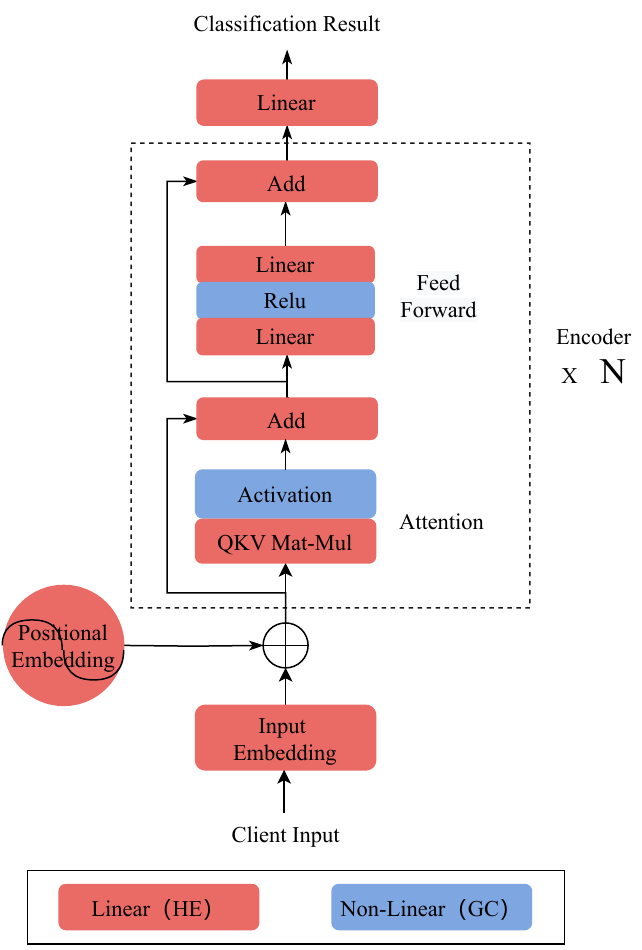}}
\caption{A transformer-based text classification model divided into HE and GC parts by operations.}
\label{model}
\end{figure}

\subsection{Base CipherFormer Protocol}\label{baseline}

Figure~\ref{model} shows a typical transformer-based text classifier.

\textbf{Linear part.} The input embedding layer, position embedding layer, matrix linear transformation in the attention mechanism layer, linear layer in the feed-forward neural network, and finally the linear layer of the classifier are categorized into a linear type, marked in red. These layers predominantly involve multiplications and additions between plaintext matrices and ciphertext vectors. Consequently, we can effectively utilize PAHE \cite{DBLP:conf/uss/JuvekarVC18} directly for these operations.

Multiplications involving two ciphertext matrices, denoted as $Q$ and $K$, necessitate specific considerations. While it is feasible to process them using homomorphic multiplication, we are cautious about potential drawbacks such as high latency and quick noise growth. As a result, we introduce a novel interaction protocol utilizing PAHE to execute ciphertext matrix multiplication in 2 rounds, as depicted in Algorithm~\ref{mul}. In the algorithm, $\mathit{S}$ denotes the server, and $\mathit{C}$ denotes the client. $[X]$ denotes the ciphertext of $X$ encrypted by private-key homomorphically.

\begin{algorithm}[htb] 
\caption{ Ciphertext Matrix Multiplication } 
\label{mul} 
\begin{algorithmic}[1] %
\REQUIRE \  

$\mathit{S}$: cipertext $[X]$, $[Y]$, random plaintext matrices ${R_1}$, ${R_2}$;\\
$\mathit{C}$: private-key.
\ENSURE  \ 

$\mathit{S}$: cipertext $[XY]$.
\STATE Server adds ${R_1}$, ${R_2}$ to cipertext $[X]$, $[Y]$ homomorphically.
\STATE Server sends $[X-{R_1}]$, $[Y-{R_2}]$ to client.
\STATE Client receives and decrypts $[X-{R_1}]$ and $[Y-{R_2}]$, computes $(X-{R_1})(Y-{R_2})$ in plaintext space and encrypts the result.
\STATE Client encrypts $X-{R_1}$, $Y-{R_2}$ using the \textit{diagonal approach} to get $[X-{R_1}]^{\prime}$, $[Y-{R_2}]^{\prime}$.
\STATE Client sends $[X-{R_1}]^{\prime}$, $[Y-{R_2}]^{\prime}$, $[(X-{R_1})(Y-{R_2})]$ to server.
\STATE Server invokes $\Pi_{SIMDScMult}$ to compute ${R_1}[Y-{R_2}]^{\prime}$ and $[X-{R_1}]^{\prime}{R_2}$, computes $RS$.
\STATE Server adds $RS$, ${R_1}[Y-{R_2}]^{\prime}$ and $[X-{R_1}]^{\prime}{R_2}$ to $[(X-{R_1})(Y-{R_2})]$ homomorphically.
\end{algorithmic}
\end{algorithm}

In Algorithm~\ref{mul}, a clever technique is employed to streamline the homomorphic computation overhead by repacking masked plaintexts. It is noteworthy that upon the server obtains the $Q$ and $K$ matrices, they are already the result of a \textit{hybrid approach} (for a comprehensive understanding of homomorphic matrix-vector multiplication approach, we recommend referring to \cite{DBLP:conf/crypto/HaleviS14, DBLP:conf/uss/JuvekarVC18} for detailed insights). This implies that the matrix $Q$ or $K$ comprises $n$ ciphertext vectors, where $n$ denotes the number of rows in the matrix. Continuing the multiplication of $Q$ and $K^T$ would necessitate performing homomorphic matrix-vector multiplication $2n^2$ times, yielding $n^2$ ciphertext vectors, with useful information confined solely to the first slot.

To mitigate this, the \textit{diagonal approach} \cite{DBLP:conf/crypto/HaleviS14} is employed in step 4. This step involves repacking and encrypting the plaintext matrices $X-{R_1}$ and $Y-{R_2}$, enabling the server to leverage the \textit{hybrid approach} in step 6 to execute only $2n$ homomorphic matrix-vector multiplications. Consequently, the number of ciphertext vectors storing the final result of the matrix multiplication is reduced from $n^2$ to $n$.

Algorithm~\ref{mul} ensures the security of private model inference. The random sampling of ${R_1}$ and ${R_2}$ in the plaintext space ensures indistinguishability between the obtained values after client decryption ($X-{R_1}$ and $Y-{R_2}$) and a random uniform distribution. Consequently, the client is incapable of extracting any weight information from the model. On the server side, as it lacks the private-key for the HE scheme, decryption of any ciphertext is unattainable. Therefore, the server is also unable to access the plaintext information of any intermediate computation output.


\textbf{Non-linear part.} The $ReLU$ function in the feed-forward neural network and the $Softmax$ function in the attention mechanism fall into the non-linear type due to the comparison and exponential operations they involve, marked in blue in Figure~\ref{model}. We utilize GC to handle these nonlinear operations effectively. 

The solution for the $ReLU$ function can be found in \cite{DBLP:conf/uss/JuvekarVC18}. However, to the best of our knowledge, no GC-friendly solution has been proposed for the $Softmax$ function in the attention mechanism layer. For $Softmax$ classifiers in CNN models, the study \cite{DBLP:conf/dac/RouhaniRK18} concentrates solely on the largest element in the output vector, employing GC with comparators and multiplexers. Unfortunately, this approach is not applicable to the attention mechanism. Additionally, the studies \cite{DBLP:conf/sp/MohasselZ17, sci3} suggest a substitution of the $e^x$ term in original $Softmax$ \eqref{Softmax} with $ReLU(x)$ for classification, resulting in a simplified version \eqref{relumax}. Here, we employ bold lowercase letters to represent vectors, and $\mathbf{x}[i]$ denotes the i-th element of the vector $\mathbf{x}$.
\begin{equation}
Softmax_{ori.}\left(\mathbf{x}\right)[i]=\frac{e^{\mathbf{x}[i]}}{\sum_{j=0}^{n-1} e^{\mathbf{x}[j]}}\label{Softmax}
\end{equation}

\begin{equation}
Softmax_{sim.}\left(\mathbf{x}\right)[i]=\frac{ReLU({\mathbf{x}[i]})}{\sum_{j=0}^{n-1} ReLU({\mathbf{x}[j]})}\label{relumax}
\end{equation}

When extending this scheme to attention mechanisms, we encounter unique challenges. In previous works, the $Softmax$ function was used at the final layer, invoked only once during the inference phase. The client's concern was primarily the order of elements in the output vector. However, in the attention mechanism, the $Softmax$ output serves as an input to the next layer and is invoked multiple times. This necessitates a more precise and efficient evaluation of the result.

\begin{algorithm}[htb] 
\caption{ GC design for Activation in Attention Mechanism } 
\label{reludiv} 
\begin{algorithmic}[1] %
\REQUIRE \  

$\mathit{S}$: additive shares $\mathbf{s_x}=\mathbf{r}$, $\mathbf{s_y}=\mathbf{r^\prime}$;\\
$\mathit{C}$: additive shares $\mathbf{c_x}=\mathbf{x-r}$;\\
$\mathit{PUBLIC}$: fixed-point number bitwidth $w$, fractional part bitwidth $f$.
\ENSURE  \ 

   $\mathit{C}$: $\mathbf{c_y}=\mathbf{y-r^\prime}$, \\where $\mathbf{y}[i]={ReLU(\mathbf{x}[i])}/{\sum_{j=0}^{n-1} ReLU({\mathbf{x}[j]})}$.
\STATE $in \gets p/2$;

\FOR{$i=0$ to $n-1$} 
    \STATE $\mathbf{x}[i] \gets A2GCircuit(\mathbf{c_x}[i], \mathbf{s_x}[i])$;
    \STATE $\mathbf{c_x^{scale}}[i], \mathbf{s_x^{scale}}[i] \gets RShiftCircuit(\mathbf{c_x}[i], \mathbf{s_x}[i], f)$;
    \STATE $\mathbf{x^{scale}}[i] \gets A2GCircuit(\mathbf{c_x^{scale}}[i], \mathbf{s_x^{scale}}[i])$;
    \STATE $sum \gets ADDCircuit(in, \mathbf{x^{scale}}[i])$;
    \STATE $in \gets sum$;
\ENDFOR \textcolor{blue}{\COMMENT{sum $ReLU(\mathbf{x}[j])$}}

\FOR{$i=0$ to $n-1$}
    \FOR{$j=w-1$ to $f-2$}
        \STATE $\mathbf{y}[i][j] \gets fixedZeroWire()$;
    \ENDFOR
    \STATE $compare \gets LShiftCircuit(sum, f-1)$;
    \FOR{$j=f-1$ to $0$}
        \STATE $tmp, nonneg \gets SUBCircuit(\mathbf{x}[i], compare)$;
        \STATE $\mathbf{y}[i][j] \gets nonneg$;
        \STATE $\mathbf{x}[i] \gets MUXCircuit(\mathbf{x}[i], tmp, nonneg)$;
        \STATE $compare \gets RShiftCircuit(compare, 1)$;
    \ENDFOR \textcolor{blue}{\COMMENT{division}}
    \STATE $\mathbf{c_y}[i] \gets G2ACircuit(\mathbf{y}[i], \mathbf{s_y}[i])$;
\ENDFOR
\end{algorithmic}
\end{algorithm}

Consequently, we have tailored the garbled circuit that computes $Softmax_{sim.}\left(\mathbf{x}\right)$ in the attention mechanism. Algorithm~\ref{reludiv} illustrates the GC design for the activation function. Here, we expand the notation of $[]$: $\mathbf{x}[i][j]$ signifies the j-th bit of the i-th element of the vector $\mathbf{x}$. The input to the algorithm is additive shares $\mathbf{s_x}$, $\mathbf{s_y}$ of vector $\mathbf{x}$, and random share $\mathbf{s_y}$ generated by server. The output is the additive share $\mathbf{c_y}$ of vector $\mathbf{y}$ on the client side, which, when added to the server's prepared share $\mathbf{s_y}$, is precisely equal to vector $\mathbf{y}$. In this pseudo-code, $A2GCircuit$ and $G2ACircuits$ denote switching circuits between the additive share and corresponding GC representation. The notation $\mathbf{c_x^{scale}}, \mathbf{s_x^{scale}} \gets RShiftCircuit(\mathbf{c_x}[i], \mathbf{s_x}[i], f)$ signifies the right-shifting of $\mathbf{c_x}[i], \mathbf{s_x}[i]$ by $f$ bits, with the outputs indicated by $\mathbf{c_x^{scale}}[i], \mathbf{s_x^{scale}}[i]$. The adders, subtractors, and multiplexers are straightforward in their functionality, as indicated in the pseudo-code. It is important to note that GC is created before evaluation, so ``$\gets$" does not imply an assignment but rather signifies the declaration of the circuit's output wire indices.

We optimize the sum and division operations based on the characteristics of fixed-point arithmetic. In transformer-based models, the vector size $n$ can be large (e.g., 100), implying that the dividend ${\sum_{j=0}^{n-1} ReLU({\mathbf{x}[j]})}$ may also be large. To prevent overflow and maintain precision, we replace the left shift of the quotient with a corresponding right shift of the dividend, summing the scaled $ReLU({\mathbf{x}[j]})$ values to obtain the dividend. In division, we observe that only the least significant $f$ bits of the quotient (i.e., the fractional part) may not be 0. Therefore, we directly connect other output bits to the 0-wire without actually computing them. Our specialized GC divider significantly reduces the computational complexity compared to a regular divider.

\subsection{Optimization 1: Lightweight Attention Mechanisms} \label{optone}

The complexity of attention mechanisms poses a challenge in private model inference, prompting an exploration of finding lightweight attention mechanisms at the cost of a slight accuracy decrease. A discussion on the role of $Softmax$ attention \cite{DBLP:conf/iclr/QinSDLWLYKZ22} provides insight into this quest. It states that providing non-negativity of attention is the crucial feature of $Softmax$ function, compared to other features such as reweighting and normalization.


Based on the findings in \cite{DBLP:conf/iclr/QinSDLWLYKZ22}, we adapt the attention mechanism from \eqref{atten1} to \eqref{atten2} for improved efficiency without a substantial sacrifice in accuracy. This adaptation involves eliminating the normalization feature of the $Softmax_{sim.}$ function by removing the division operation. Therefore, this modification allows a significant reduction of GC gates. Nevertheless, the remaining $ReLU$ function preserves the crucial non-negativity feature.
\begin{equation}
Attention(Q, K, V)=Softmax_{sim.}\left(\frac{Q K^{T}}{\sqrt{d_{k}}}\right) V\label{atten1}
\end{equation}
\begin{equation}
Attention(Q, K, V)=ReLU\left(\frac{Q K^{T}}{\sqrt{d_{k}}}\right) V\label{atten2}
\end{equation}

Furthermore, the order of association in the $QKV$ matrix multiplication can be adjusted to further minimize the computational overhead of the HE part, which introduces subtle differences in the calculation process. Modify \eqref{atten2} as \eqref{atten3}:
\begin{equation}
Attention(Q, K, V)=\frac{1}{\sqrt{d_k}} {ReLU}(Q) {ReLU}(K^T) V \label{atten3}
\end{equation}

The reduction in computational overhead stems from the characteristics of the text classification task. In the matrices $Q$ and $K$, the row numbers correspond to the text sequence length $L$, while the column numbers represent the feature dimension of a single word, denoted as $d$. Typically, $L \gg d$ holds. As a result, the complexity of matrix multiplication is diminished from $O(L^{2}d)$ to $O(Ld^2)$, leading to a decrease in homomorphic matrix-vector multiplication invocations in Algorithm~\ref{mul} from $2L$ times to $L+d$ times. Table~\ref{opt1table} presents the summarized improvements between the baseline ``CipherFormer" and this optimized version, denoted as ``Opt. 1".

\begin{table}[htbp] 
    \begin{center}
    \caption{Comparison of overhead between Baseline and Optimization 1 involves measuring the numbers of non-XOR and XOR GC gates for the activation at an input size of 100, bitwidth of 20, and fractional part bitwidth of 9. The notation $L \gg d$ indicates that $L$ significantly exceeds $d$ in the $\mathcal{O}$ complexity for homomorphic computation.}
    \begin{tabular}{@{}ccccc@{}}
    \toprule
    Scheme   & \#Non-XOR & \#XOR  & SIMDScMult & Matrix Mult \\ \midrule
    CiperFormer & 56.3k     & 174.8k & $O(2L)$         & $O(L^2d)$             \\
    +Opt. 1   & 16.2k     & 51.4k  & $O(L+d)$        & $O(Ld^2)$             \\ \bottomrule
    \end{tabular}
    \label{opt1table}
    \end{center}
\end{table}

\subsection{Optimization 2: Mixed-bitwidths} \label{opttwo}

To minimize the latency of non-linear function inference, advanced mixed-bitwidth techniques are employed. SiRNN \cite{DBLP:conf/sp/RatheeRGGSCR21} noted that properly reducing the bitwidth of non-linear activations has only a marginal impact on inference accuracy. Their protocol is constructed using SS and OT over the $Z_{2^k}$ ring. Given these insights, the application of mixed-bitwidth techniques to our HE-GC protocol is a natural expectation.

We implemented the bitwidth switching protocol for the HE-GC version. In practice, the implementation in garbled circuits is relatively straightforward compared to the protocol that switches between a larger ring $Z_{2^m}$ and a smaller ring $Z_{2^n}$ in \cite{DBLP:conf/sp/RatheeRGGSCR21}. In GC, fixed-point numbers are represented in bits, with each bit assigned to a wire. Therefore, we can simply fix the decimal point and then remove the wires for the corresponding least and most significant bits to reduce the bitwidth. Similarly, by padding 0-wire in the least and most significant bits of the output, we can extend the bitwidth. This optimization is denoted as ``Opt. 2".

\section{Experiments}

\subsection{Experimental Setup}

\textbf{Environment}. We measure the accuracy, computational latency, and communication overhead for private model inference of transformer models on an Ubuntu 16.04 TLS virtual machine with Linux version 4.15.0-142-generic and gcc version 5.4.0. The memory size of the VM is 7.5G. We use two threads on the VM to simulate the server and client. 

\textbf{Cryptographic Primitives}.
Both HE and GC utilize 128-bit security parameters in our implementation. We leverage the HE library described in \cite{DBLP:conf/uss/JuvekarVC18}. For the GC, we employ the open-source code JustGarble, which offers fundamental gate circuit primitives along with optimizations like half-gate techniques \cite{DBLP:conf/eurocrypt/ZahurRE15}.

\textbf{Model and datasets}. 
We evaluate the inference accuracy on text classification datasets using the transformer model with the structure depicted in Figure~\ref{model}. Our experiments cover four datasets: the 10-class dataset Yahoo! Answers, the news topic 4-class dataset AG\_News, and the sentiment 2-class datasets IMDB and Yelp. In the AG\_News dataset, the maximum sentence word counts $L$ is set to 100, and the feature dimension $d$ after embedding is 32. For the Yelp, IMDB, and Yahoo datasets, we use $L=128$ and $d=64$ due to the larger dataset size. The encoder number N is 1 in AG\_News, IMDB, and Yelp datasets. N = 2 in Yahoo! Answers datasets. 

\subsection{Mixed-bitwidths Performance}

The adjustment of the model structure and the incorporation of mixed bitwidths typically result in a reduction of private model inference accuracy, introducing a trade-off between model accuracy and evaluation overhead.


We begin the discussion of this trade-off by assessing the effectiveness of the mixed-bitwidth strategy in the CipherFormer Protocol. The results presented in Table~\ref{mixedtable} showcase the private model inference accuracy on the AG\_NEWS dataset. ``Original Bitwidth" refers to a 20-bit fixed-point bitwidth with a 9-bit fractional part bitwidth. ``Low Bitwidth" indicates a scenario where all fixed-point parameters bitwidth in our model are reduced to 16 bits, with 7-bit fractional part bitwidth. ``Mixed Bitwidth" signifies that only the non-linear activation parameters are reduced to a 16-bit fixed-point bitwidth, while the rest remain at a 20-bit bitwidth. The experimental results indicate that in our HE-GC scenario, a simple reduction in fixed-point bitwidth significantly diminishes accuracy, whereas mixed-bitwidth strategy incurs almost no loss of accuracy, consistent with SiRNN \cite{DBLP:conf/sp/RatheeRGGSCR21}.

\begin{table}[htbp]
    \begin{center}
        \caption{Comparison of Different Bitwidth Strategies.}
\begin{tabular}{@{}cccc@{}}
\toprule
\multirow{2}*{Scheme}      & \multicolumn{3}{c}{Accuracy(\%)} \\
~ & Original Bitwidth & Low Bitwidth & Mixed bitwidth \\ \midrule
CiperFormer & 87.70             & 58.46        & 87.73          \\
+Opt. 1     & 90.02             & 77.57        & 90.02          \\ \bottomrule
\end{tabular}
    \label{mixedtable}
    \end{center}
\end{table}

\begin{figure}[htbp]
\centerline{\includegraphics[width=0.4\textwidth]{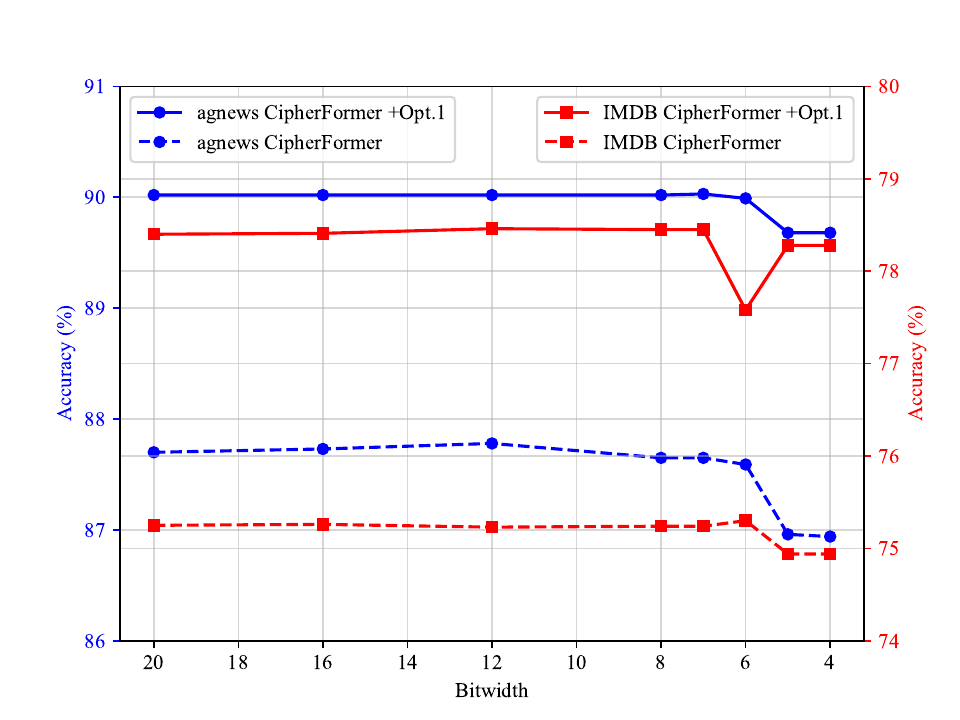}}
\caption{The accuracy in different activation function bitwidth strategy over two datasets AG\_NEWS and IMDB.}
\label{bitwidthpic}
\end{figure}

Furthermore, we introduce additional datasets to test how activation function bitwidth in mixed-bitwidth strategy affects inference accuracy, illustrated in Figure~\ref{bitwidthpic}. In this test, we proportionally decrease the fractional part bitwidth as the activation bitwidth decreases. The figure demonstrates that reducing the activation bitwidth to 8 bits maintains model accuracy, while further decreases result in accuracy loss. Consequently, we set the mixed bitwidth to 8 for the activation function and 20 for other parameters in Opt. 2.

\subsection{Private Inference Accuracy}
We measure the private inference accuracy of both the CipherFormer protocol and its two optimizations across three datasets, comparing them with the accuracy of plaintext floating-point inference (see Table~\ref{Accuracy}). We add Opt.1 and Opt.2 to the baseline CipherFormer serially. The experimental results demonstrate our ability to protect privacy during inference without causing a significant decrease in accuracy.

\begin{table}[htbp]
    \begin{center}
        \caption{Comparison of Inference Accuracy.}
\begin{tabular}{@{}ccccc@{}}
\toprule
\multirow{2}*{Datasets}   &  \multicolumn{4}{c}{Accuracy(\%)}          \\
        ~                & Plaintext & CipherFomer & +Opt.1 & +Opt.2 \\ \midrule
AG\_News                 & 90.57     & 87.70       & 90.02  & 90.03  \\
IMDB                     & 83.46     & 75.25       & 78.40  & 78.45   \\
Yelp                     & 91.73     & 91.80       & 91.79  & 91.79  \\ \bottomrule
\end{tabular}
    \label{Accuracy}
    \end{center}
\end{table}

\subsection{Latency and Communication Overhead}

We evaluate the latency and communication overhead of private inference on the Yelp Review dataset, as detailed in Table~\ref{overhead}. Offline latency involves activities such as HE key distribution, garbled circuit generation, and model weight matrix encoding. Online latency involves homomorphic evaluation, OT sub-protocol in GC, garbled circuit evaluation, and other related activities. Our proposed optimizations yield a 1.23x-1.55x speedup in offline settings and a 1.23x-1.47x speedup in online inference, reducing communication overhead by 40\%, compared with the baseline. This comes at a reasonable cost of a slight accuracy decrease, as recalled in Table~\ref{Accuracy}. 

\begin{table}[htbp]
    \begin{center}
        \caption{Comparison of latency and communication overhead.}
\begin{tabular}{@{}ccccc@{}}
\toprule

\multirow{2}{*}{Scheme} & \multicolumn{2}{c}{Latency(s)} & \multicolumn{2}{c}{Comm.(MB)} \\
             & Offline & Online & Offline & Online \\ \midrule
CipherFormer & 21.89   & 7.59   & 366.4   & 54.0   \\
+Opt.1       & 17.75   & 6.16   & 101.9   & 46.0   \\
+Opt.2       & 14.07   & 5.15   & 42.5    & 32.0   \\ \bottomrule
\end{tabular}
    \label{overhead}
    \end{center}
\end{table}



\subsection{Comparison with prior work}

We conduct a comparison between the optimized CipherFormer and the advanced HErBERT \cite{comi2021herbert} on two identical text classification tasks, as detailed in Table~\ref{Parameters of the 3-segment continuum robot}. 
Our inference accuracy surpasses HErBERT by 3\% to 11\%, attributed to our customized GC-friendly attention mechanism protocol, rather than simply discarding nonlinear terms as done in HErBERT. Furthermore, despite the larger feature dimension and encoder number employed in our transformer model compared to HErBERT, our more complex model achieves inference speeds 7.7x-11.9x faster than HErBERT, demonstrating the efficiency of CipherFormer and its optimizations.

\begin{table}[htbp]
    \caption{Comparison of CipherFormer and HErBERT.}
	\label{Parameters of the 3-segment continuum robot}
	\begin{center}
	\begin{threeparttable}  
	\begin{tabular}{@{}ccccc@{}}
	\toprule

\multirow{2}{*}{Scheme} & \multicolumn{2}{c}{Accuracy(\%)} & \multicolumn{2}{c}{Latency(s)} \\
 & Yelp & Yahoo! & Yelp & Yahoo! \\ \midrule
CipherFormer\tnote{*} & 91.79 & 70.47 & 5.15 & 9.21 \\
HErBERT & 88.2 & 59.26 & 40 & 110\\ \bottomrule
	\end{tabular}
	\begin{tablenotes}    
        \footnotesize               
        
             \item[]  *with 2 optimizations in \S~\ref{optone} and \S~\ref{opttwo}.
      \end{tablenotes}            
    \end{threeparttable}      
	\end{center}
\end{table}


\section{Conclusion}
We propose CipherFormer, a transformer private model inference protocol that employs both HE and GC with a focus on minimizing communication rounds. Our protocol includes carefully designed procedures for cipher matrix multiplication in HE and specialized GC algorithms for activations in the attention mechanism. Through optimization, we achieve a 32\% reduction in latency while preserving accuracy. This work addresses shortcomings in prior approaches, offering an efficient and accurate solution for transformer model inference in specific scenarios. While our approach demonstrates promising results, validating its effectiveness on more intricate transformer models across diverse tasks remains an open avenue for further exploration.


\bibliographystyle{IEEEtran}
\bibliography{IEEEabrv,mylib}

\end{document}